\documentclass{article}
\usepackage{PRIMEarxiv}
\usepackage[utf8]{inputenc} 
\usepackage[T1]{fontenc}    
\usepackage{hyperref}       
\usepackage{url}            
\usepackage{booktabs}       
\usepackage{amsfonts}       
\usepackage{nicefrac}       
\usepackage{microtype}      
\usepackage{lipsum}
\usepackage{fancyhdr}       
\usepackage{graphicx}       
\usepackage{subcaption}
\usepackage{amsmath}
\usepackage{arydshln}

\begin{document}

\title{Fast Node Vector Distance Computations using Laplacian Solvers}

\author{Michele Coscia\\
IT University of Copenhagen\\
Copenhagen, Denmark\\
\texttt{mcos@itu.dk}\\
\And
Karel Devriendt\\
Max Planck Institute for Mathematics in the Sciences\\
Leipzig, Germany\\
\texttt{karel.devriendt@mis.mpg.de}
}

\maketitle

\begin{abstract}
Complex networks are a useful tool to investigate various phenomena in social science, economics, and logistics. Node Vector Distance (NVD) is an emerging set of techniques allowing us to estimate the distance and correlation between variables defined on the nodes of a network. One drawback of NVD is its high computational complexity. Here we show that a subset of NVD techniques, the ones calculating the Generalized Euclidean measure on networks, can be efficiently tackled with Laplacian solvers. In experiments, we show that this provides a significant runtime speedup with negligible approximation errors, which opens the possibility to scale the techniques to large networks.
\end{abstract}

\section{Introduction}
Complex networks are useful for a number of tasks. One prominent example is tracking the propagation of a phenomenon through a complex system. Examples range from diseases \cite{bajardi2011human, kraemer2020effect, vespignani2020modelling}, memes/behaviors \cite{granovetter1978threshold, weng2014predicting, friggeri2014rumor, coscia2022posts, hohmann2023quantifying}, or product adoption \cite{leskovec2007dynamics,watts2007viral} through a social network; productive knowledge in international trade \cite{hausmann2014atlas, tacchella2012new}; or goods in network modeling problems in logistics \cite{pele2009fast}.

The Node Vector Distance (NVD) term has been recently used to group these tasks under a common structure \cite{coscia2020node}. In NVD, the phenomenon is represented as a vector recording one value per node. Then two vectors from different phenomena, or from the same phenomenon at different observation times, can be compared. Specifically, with NVD one can calculate their distance, network variance, or correlation.

Most useful NVD techniques share a drawback: they are computationally complex to calculate. This severely limits their practical applicability to nodes containing a handful thousands of nodes, a far cry from the (tens or hundreds) million of nodes of the most interesting complex networks.

In this paper we focus specifically on those NVD techniques based on the inversion of the graph Laplacian \cite{coscia2020generalized}. We do so for two reasons. First, these measures are among the most intuitive available. Second, because it turns out that the most computationally intensive part of calculating such measures is not necessary.

We show how the already existing collection of techniques known as ``Laplacian solvers'' \cite{spielman2004nearly} can be directly applied to the Generalized Euclidean NVD technique, greatly reducing its computational complexity and allowing the analysis of really large complex networks.

In our experiments we show how much runtime we gain in synthetic networks of growing sizes, showing an empirical estimation of the new computational complexity. We also do a brief analysis of the memory consumption. Finally, we show the practical applicability on a number of real world networks. The latter experiments also shows that, even if Laplacian solvers do not provide exact solutions, the approximation they induce is negligible for all practical purposes.

All the experiments we run can be reproduced with the material we provide\footnote{\url{https://www.michelecoscia.com/?page\_id=1733\#nvdfast}}.

\section{Related Works}

\subsection{Node Vector Distance}
Node Vector Distance (NVD) is a collection of techniques to estimate the network distance between node vectors -- vectors recording one value per node \cite{coscia2020node}. NVD has a number of applications in network science, it can be used to track disease spreading \cite{bajardi2011human}, estimate the complexity of a country's economy \cite{hausmann2014atlas}, or quantify ideological polarization on social media \cite{hohmann2023quantifying}. The techniques at the basis of NVD can also be used to estimate how dispersed a variable is in a network \cite{devriendt2022variance}, as well as calculating the correlations between node vectors on a network \cite{coscia2021pearson}.

There are a number of different approaches one can take. One can apply graph signal processing techniques via the graph Fourier transform \cite{shuman2013emerging, shuman2016vertex}. Another popular approach is to compute the optimal way to transport the weights of one vector to another with respect to the distance in the network, giving rise to the Earth Mover Distance \cite{pele2009fast, zhao2019moverscore}.

In this paper we focus on a different class of solutions, which we label ``Generalized Euclidean''. In this class, one adapts the classical Euclidean distance to the graph setting. In the case of regular Euclidean distance, the node vectors are embedded in a space where all dimensions contribute equally -- here, the distance is induced by the inner product represented by the identity matrix, so there is no distinction between the nodes. In the case of Generalized Euclidean distance, the node vectors are embedded in a complex space represented by the graph; more precisely, the distance between node vectors in this space is given by computing the quadratic product of their difference with the pseudoinverse Laplacian matrix as in Section 3.1. Note that using the pseudoinverse Laplacian is not a unique solution, as there are other ways to take into account the graph structure in the Euclidean formula \cite{coscia2020generalized}.

Since the pseudoinverse Laplacian is the technique we focus on in this paper, we will provide more details about this approach in Section \ref{sec:meth}. For the purpose of this section, we only need to mention that pseudoinverting the Laplacian is computationally complex, but not necessary. One can achieve an approximate result by using Laplacian solvers, which we discuss now.

\subsection{Laplacian Solvers}\label{sec:related-solvers}
Laplacian solvers are a class of solutions to problems in the form $Lx = b$, where $L$ is the Laplacian of an undirected
graph \cite{vishnoi2013lx}. These solvers have a number of applications in graph partitioning and specification.

Laplacian solvers make use of a number of techniques to solve the $Lx = b$ problem in near linear time \cite{spielman2004nearly, spielman2011spectral, spielman2013local, spielman2014nearly}. Examples include sparse approximate Gaussian elimination \cite{kyng2016approximate}, building a chain of progressively sparser graphs \cite{koutis2011nearly, koutis2014approaching}, and recursive graph preconditioning \cite{spielman2014nearly}.

One major drawback of the methods cited so far is that they only work with undirected graphs. However, there is a collection of techniques that work on directed graphs as well \cite{cohen2016faster, cohen2017almost}.

\section{Methods}\label{sec:meth}

\subsection{Generalized Euclidean}
Let us assume we are working with a graph $G = (V, E)$, with $V$ being the set of nodes and $E \subseteq V \times V$ the set of edges -- pairs of nodes. For this paper we assume to work with undirected graphs: if $u,v \in V$ and $(u,v) \in E$, then $(u,v) = (v,u)$. The graphs can be weighted, i.e. each edge can have a positive real weight $w > 0$ -- although, in this paper, we ignore weights (including them does not change any of our conclusions).

We can define a number of useful matrices. $A$ is the adjacency matrix of $G$, with $A_{uv} = 1$ if $(u,v) \in E$ and $A_{uv} = 0$ otherwise. $D$ is the degree matrix, the degree being the number of connections a node has. $D$ contains the degree of a node in the main diagonal and zero elsewhere. The Laplacian matrix is defined as $L = D - A$, i.e. it contains the degree of a node on the main diagonal and $L_{uv} = -1$ if $(u,v) \in E$.

The Laplacian is useful to solve a number of problems. For instance, it can be used to solve the discrete heat exchange problem. If $h$ contains the heat value for each node of the network, we can use the Laplacian to estimate how heat propagates through the graph. This is done by solving the differential equation $\dfrac{\partial h}{\partial t} = -Lh$ \cite{coifman2006diffusion}. It can also be used for spectral clustering \cite{von2007tutorial}.

It follows that the Laplacian is helpful to understand the relationships between nodes. Previous work has exploited this fact to use the Laplacian as the matrix defining the space in which a Generalized Euclidean (GE) distance measure lives. If we are given two vectors $a$ and $b$, each with $|V|$ entries, then their network distance is:

$$ \delta_{G,a,b} = \sqrt{(a - b)^TL^{\dagger}(a - b)}.$$

where $L^\dagger$ is the (Moore--Penrose) pseudoinverse of $L$. $L$ cannot be inverted directly, because it is singular. To calculate $L^\dagger$ one needs to perform a singular value decomposition (SVD) of $L$. Herein lies the main issue with this measure: SVD requires $\mathcal{O}(|V|^\alpha)$ time to be solved, with $\alpha$ larger than $2$ and smaller than $3$. This makes GE intractable for all but trivially sized graphs.

\subsection{Laplacian Solvers}
A Laplacian solver is a technique that is able to solve systems of linear equations in the form of $Lx = b$ in near linear time. To explain each Laplacian solver techniques in depth goes beyond the scope of this paper. In Section \ref{sec:related-solvers} we provide further references. In this section, we briefly mention how some of these solvers work.

Sparse approximate Gaussian elimination \cite{kyng2016approximate} works by performing an approximate sparse Cholesky decomposition. The Cholesky decomposition is an efficient algorithm for solving systems of linear equations. The issue is that, locally, $L$ does not satisfy the sparsity assumption for the Cholesky decomposition -- e.g. in case of large cliques. Thus, cliques need to be sampled and then the regular Cholesky decomposition can be applied.

Spectral graph sparsification \cite{koutis2011nearly, koutis2014approaching} works by taking $G$ and sparsify it to $G'$ in such a way that $G$ and $G'$ have very similar spectra. This is done iteratively via a preconditioning chain. After the first spectral sparsification, $G'$ is then contracted by eliminating nodes of degree $1$ and $2$. This can be done efficiently, because the spectrum of the Laplacian is related to the cut problem, and it is possible to sparsify the graph while preserving its cuts.

Recursive graph preconditioning \cite{spielman2014nearly} puts together the previous two approaches by recursively sparsifying $G$ via a partial Cholesky factorization, ensuring a low condition number at every step in the recursion.

\section{Experiments}

\subsection{Setup Details}

\subsubsection{Implementation}
We implement the GE function in Julia (version 1.8.0). We use the Laplacians.jl package\footnote{\url{https://danspielman.github.io/Laplacians.jl/dev/}} for Julia to access implementations of the Laplacian solvers (version 1.3.0). We use the methods' names provided in the package to refer to the various methods we use here. We run our code on a Intel Xeon Platinum 8358 at 2.60GHz. 

\subsubsection{Synthetic Data}
We test the Laplacian solvers on a number of synthetic networks, which allow us to vary both the number of nodes $|V|$ and the graph's density by changing the average degree. We use different models because each of them can reproduce some of the common properties we find in real world networks. Specifically, we use:

\begin{itemize}
\item Erd\H{o}s--R\'{e}nyi (ER): this is a $G_{n,m}$ model where we create a graph with $|V| = n$ and $|E| = m$. The edges are assigned uniformly at random by extracting two random node ids. This network model reproduces well the small world feature of real networks -- the resulting networks have small diameters \cite{coscia2021atlas}.
\item Barab\'{a}si-Albert (BA): we grow this network by adding one node at a time. Each node connects to $k$ already existing nodes, with $k$ being a parameter. Existing nodes receive new connections with a probability directly proportional to their degree. This network reproduces well both small world property and broad degree distributions \cite{barabasi1999emergence}.
\item Watts-Strogatz (WS): this network starts from a circle graph where nodes are connected to all of their $k$ closest neighbors. Then, each edge is rewired randomly with probability $p$, with $k$ and $p$ as parameters. This model reproduces the high clustering and small world features \cite{watts1998collective}.
\item Stochastic Blockmodel (SBM): in this model, as an input, the user partitions nodes into groups and specifies two probabilities. $p_{in}$ determines the probability of connecting to a node inside the same group, and $p_{out}$ regulates the connections to nodes outside the group. This model can generate network communities \cite{holland1983stochastic}.
\end{itemize}

For all models, we make sure to directly compare networks with roughly the same number of edges.

\subsubsection{Real World Data}\label{sec:exp-setup-realdata}
For our applications section (Section \ref{sec:applications}) we also make use of real world data, to showcase the usefulness of GE -- and, as a consequence, the need for efficient ways to estimate it. Specifically we use:

\begin{itemize}
\item Section \ref{sec:applications-pol} (US Congress): networks from the roll call votes in the House of Representatives -- one network per congress edition --, using data from Voteview.com \cite{lewis2019voteview}. Each node is a representative and they are connected if the two representatives have co-voted on bills more often than the average same-party pair. The procedure to build these networks has been used multiple times in the literature \cite{andris2015rise, hohmann2023quantifying}.
\item Section \ref{sec:applications-runtime} (Various): networks that come from a variety of papers, retrieved via the network catalogues SNAP \cite{snapnets} and Netzschleuder \cite{peixoto2020netzschleuder}.
\end{itemize}

We use small networks in Section \ref{sec:applications-pol} because we want to show how accurate the Laplacians solvers can be in quantifying the GE values against the exact result obtained via SVD -- thus we need to be able to run SVD. The larger networks in Section \ref{sec:applications-runtime} are used to showcase the possibilities opened by the Laplacian solvers that are not available to the exact solutions via the Laplacian pseudoinverse.

\subsection{By Network Size}
In this section we test the effect of the size of the network on the running time and memory consumption of the Laplacian solvers against the baseline using the pseudoinversion via SVD. We split the size test first by increasing the number of nodes while keeping the density of the network constant, and then by keeping the number of nodes fixed but increasing the network density.

\subsubsection{Runtimes ($|V|$)}
For the runtimes, we exclude outlier runs which took more than twice the average runtime. This is done to exclude compilation time from the estimate -- this issue only affects very small input sizes where compilation could take significantly longer than running time. All plots report average runtimes over ten independent runs. The exception is Baseline, for which we make a single run for $|V| = 10^4$ and we do not run for larger $|V|$ at all, due to its excessively long runtimes.

\begin{figure}
\centering
\includegraphics[width=.66\columnwidth]{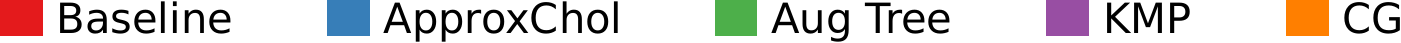}\\
\begin{subfigure}[b]{.24\columnwidth}
\centering
\includegraphics[width=\textwidth]{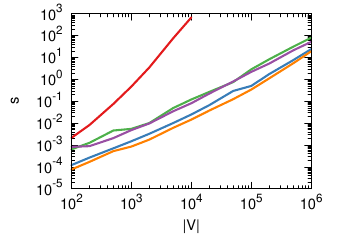}
\caption{Erd\H{o}s--R\'{e}nyi}
\end{subfigure}
\begin{subfigure}[b]{.24\columnwidth}
\centering
\includegraphics[width=\textwidth]{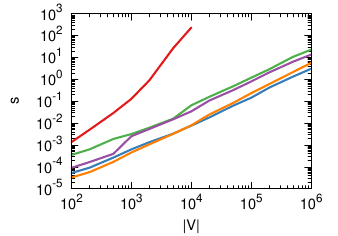}
\caption{Barab\'{a}si-Albert}
\end{subfigure}
\begin{subfigure}[b]{.24\columnwidth}
\centering
\includegraphics[width=\textwidth]{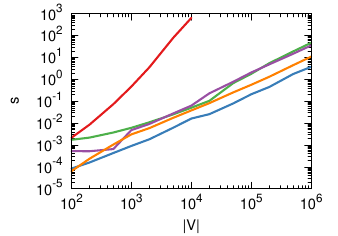}
\caption{Watts-Strogatz}
\end{subfigure}
\begin{subfigure}[b]{.24\columnwidth}
\centering
\includegraphics[width=\textwidth]{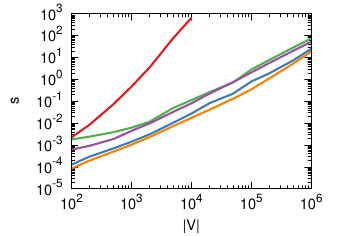}
\caption{Stochastic Blockmodel}
\end{subfigure}
\caption{The running time (y axis) against $|V|$ (x axis) for all the methods (line color) on different synthetic networks.}
\label{fig:runtime}
\end{figure}

We start by analyzing the runtimes for increasing number of nodes. Figure \ref{fig:runtime} reports the results. The first evident result is that any Laplacian solver has both a constant running time advantage and a better asymptotic complexity. Even for tiny networks of $100$ nodes, regardless of the network topology, all Laplacian solvers are at least one order of magnitude faster than the baseline. From these plots we can infer that the empirical asymptotic complexity of the baseline is $\sim \mathcal{O}(|V|^{2.6})$. For the Laplacian solvers the exact combination of solver and topology matters, but in general the empirical asymptotic complexity is between $\mathcal{O}(|V|^{1.2})$ and $\mathcal{O}(|V|^{1.4})$, in all cases decisively below $\mathcal{O}(|V|^{2})$.

The practical result is that the baseline takes at least one order of magnitude more time to compute a $|V| = 10^4$ network than any Laplacian solver takes for a network two orders of magnitude larger.

\begin{figure}
\centering
\includegraphics[width=.45\columnwidth]{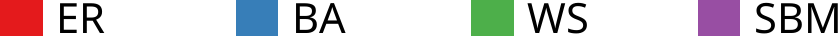}\\
\begin{subfigure}[b]{.33\columnwidth}
\centering
\includegraphics[width=\textwidth]{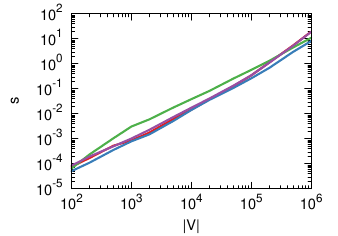}
\caption{Runtime}
\end{subfigure}
\qquad
\begin{subfigure}[b]{.33\columnwidth}
\centering
\includegraphics[width=\textwidth]{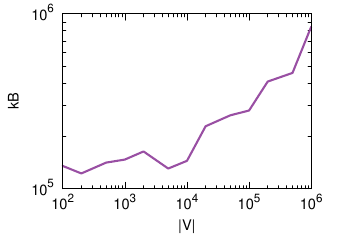}
\caption{Memory}
\end{subfigure}
\caption{Time and memory consumption (y axis) against $|V|$ (x axis) for all synthetic networks (line color) for the CG Laplacian solver.}
\label{fig:topology}
\end{figure}

Among the Laplacian solvers there is no clear overall winner. CG is the fastest for the Erd\H{o}s--R\'{e}nyi and SBM topologies, but it ties with ApproxChol for the Barab\'{a}si-Albert model, while ApproxChol is also fastest for Watts-Strogatz. The topology in general has different effects on different solvers. Picking CG as an example, Figure \ref{fig:topology}(a) shows that indeed CG runs slower for Watts-Strogatz than it does for all other topologies. Other solvers also experience different strengths and weaknesses depending on the topology -- not shown here for space issues.

\subsubsection{Runtimes (Density)}
Efficient Laplacian solvers exploit, among other things, the sparseness of a graph. It is interesting to investigate what happens to the runtime when the graphs we investigate get denser and denser. In this experiment, we fix $|V| = 10,000$, and we increase the average degree of the network from $1$ to $64$. The vast majority of real world networks have low average degrees in the single digit realm \cite{ghasemian2020stacking}, thus this domain covers the most realistic scenarios. Also in this case we report the average of ten runs, taking out outliers and ignoring compilation time. 

\begin{figure}
\includegraphics[width=.66\columnwidth]{figures/legend_methods.pdf}\\
\centering
\begin{subfigure}[b]{.24\columnwidth}
\centering
\includegraphics[width=\textwidth]{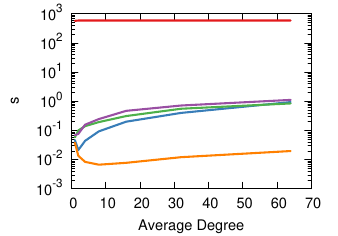}
\caption{Erd\H{o}s--R\'{e}nyi}
\end{subfigure}
\begin{subfigure}[b]{.24\columnwidth}
\centering
\includegraphics[width=\textwidth]{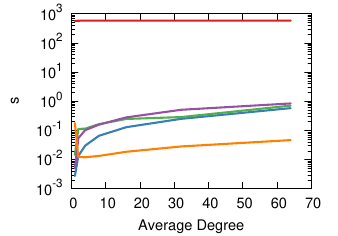}
\caption{Barab\'{a}si-Albert}
\end{subfigure}
\begin{subfigure}[b]{.24\columnwidth}
\centering
\includegraphics[width=\textwidth]{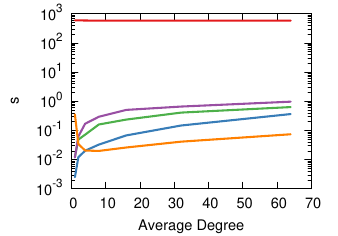}
\caption{Watts-Strogatz}
\end{subfigure}
\begin{subfigure}[b]{.24\columnwidth}
\centering
\includegraphics[width=\textwidth]{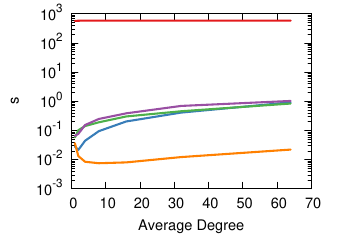}
\caption{Stochastic Blockmodel}
\end{subfigure}
\caption{The running time (y axis) against average degree (x axis) for all the methods (line color) on different synthetic networks.}
\label{fig:runtime-density}
\end{figure}

Figure \ref{fig:runtime-density} shows the results. Since the baseline works with dense matrices anyway, there is no real effect of density on its running time, which is roughly constant. Most Laplacian solvers have longer runtimes for denser networks -- as expected. All Laplacian solvers are orders of magnitude faster than the baseline, and thus represent a significant advantage.

CG shows a peculiar pattern: it takes longer for extremely sparse networks -- with average degree close to one -- then gets faster and faster for middle values of average degree between four and eight. After this, the runtimes increase with density as expected. This pattern is consistent, independently from the topology of the network. It seems that, for extremely sparse networks with average degree lower than four, CG might not be the best choice. For denser networks, however, CG can be one or two orders of magnitude faster than the other Laplacian solvers.

\subsubsection{Memory}
For the memory test we show only a single run per method, due to limitations in memory benchmarking. However, memory consumption should not be variable across runs and the results of a single run are still indicative of the overall trends. We also run a single Laplacian solver, CG, because the memory consumption for all solvers is indistinguishable in all cases.

\begin{figure}
\centering
\includegraphics[width=.2\columnwidth]{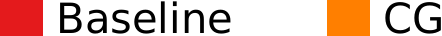}\\
\begin{subfigure}[b]{.24\columnwidth}
\centering
\includegraphics[width=\textwidth]{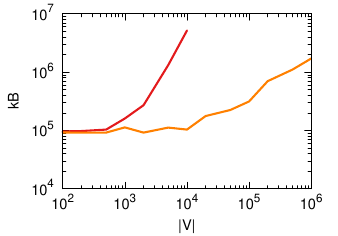}
\caption{Erd\H{o}s--R\'{e}nyi}
\end{subfigure}
\begin{subfigure}[b]{.24\columnwidth}
\centering
\includegraphics[width=\textwidth]{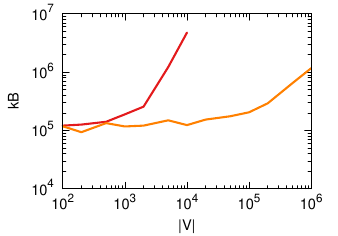}
\caption{Barab\'{a}si-Albert}
\end{subfigure}
\begin{subfigure}[b]{.24\columnwidth}
\centering
\includegraphics[width=\textwidth]{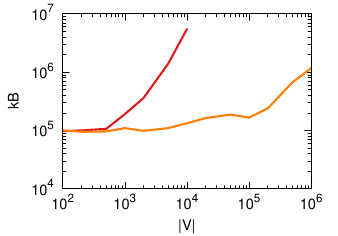}
\caption{Watts-Strogatz}
\end{subfigure}
\begin{subfigure}[b]{.24\columnwidth}
\centering
\includegraphics[width=\textwidth]{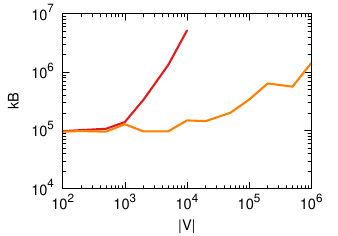}
\caption{Stochastic Blockmodel}
\end{subfigure}
\caption{Memory consumption (y axis) against $|V|$ (x axis) for all the methods (line color) on different synthetic networks.}
\label{fig:memory}
\end{figure}

From Figure \ref{fig:memory} we can see that there is a basic memory consumption coming from simply running the program. For $|V| = 10^4$, the Laplacian solvers do not add any memory consumption to this basic rate. However, the baseline needs to shift from sparse to dense matrix representations to calculate the pseudoinverse of the Laplacian. This means that its memory consumption is already between $5$ and $6$GB in our implementation even for these small networks. If we exclude the warm-up phase for $|V| < 10^3$, the memory consumption of the baseline scales exponentially.

This is not true for Laplacian solvers, here represented by CG. The total memory consumption at $|V|=10^4$ is still in the neighborhood of the basic cost of running the program. Even for $|V|=10^6$, the memory required is below $2$GB in all but one case. Asymptotically, the best function describing the growth in memory consumption by CG is linear, not exponential. Figure \ref{fig:topology}(b) does not show any significant difference in memory consumption for CG depending on the topology of the network.

\section{Applications}\label{sec:applications}

\subsection{Polarization}\label{sec:applications-pol}
The GE measure can be used to estimate polarization on social media, or any networked system where we have information about the opinions of the nodes \cite{hohmann2023quantifying}. This is done by calculating the distance between the vector recording the opinions of nodes on one side of the spectrum -- e.g. Democrats -- with the one recording the opinions of the nodes on the other side of the spectrum -- e.g. Republicans.

\begin{table}
\centering
\begin{tabular}{l|rrr}
Method & 85th & 105th & 113th \\
\hline
\hline
Baseline & 1.006 & 3.664 & 8.330\\ 
\hdashline
ApproxChol & $2.1e^{-14}$ & $2.1e^{-14}$ & $1.3e^{-13}$\\
Aug Tree & $6.9e^{-9}$ & $1.1e^{-14}$ & $5.0e^{-14}$\\
KMP & $4.4e^{-16}$ & $2.7e^{-15}$ & $1.1e^{-13}$\\
CG & $4.9e^{-10}$ & $2.7e^{-15}$ & $2.1e^{-13}$\\
\end{tabular}
\caption{The polarization scores for three US Congress networks (top row) and the difference between exact and approximate solution using a specific Laplacian solver (bottom four rows).}
\label{tab:polarization-errors}
\end{table}

For this task we use the Congress networks described in Section \ref{sec:exp-setup-realdata}. Specifically, we focus on the 85th, 105th, and 113th Congress, since they show the lowest, average, and highest value of polarization, respectively. Table \ref{tab:polarization-errors} shows the result. First, the Baseline method confirms the differences in scores between the three networks. Then we show how the four Laplacian solvers estimate the level of polarization to be practically identical to the exact one we compute via the Baseline. The largest error is in the neighborhood of $10^{-8}$, which is far below the level of precision required for such an analysis.

\subsection{Various}\label{sec:applications-runtime}
We look at a variety of networks which would all benefit from a GE analysis, in increasing size to show the speedup of the Laplacian solvers in real world scenarios. We simplify all networks to an undirected, unweighted, simple graph version, even if the original network was either directed, weighted, or multilayer. All runtimes exclude I/O operations and preprocessing, so they ignore the time it takes to read the graph from disk. Below we briefly explain what the node vectors are in each case.

\begin{itemize}
\item \textit{Hiring}: we can calculate the distance between the region in which a university is located, by analyzing the hiring patterns.
\item \textit{EUAir}: we can calculate the distance between airlines depending on which airports they serve.
\item \textit{EUCore}: we have communities based on email exchange, and GE could tell the distance between community pairs.
\item \textit{Open Flights}: we can calculate the distance between countries based on how the airlines connect their airports.
\item \textit{LastFm}: we can calculate the distance between countries -- a metadata we have about the users -- based on their friendships on the platform.
\item \textit{Wiki RFA}: we can calculate the distance between admins and non-admins in the voting network.
\item \textit{Fly Brain}: we can calculate the distance between neuron types in the neural network.
\item \textit{Twitter15m}: we can measure the distance between two hashtags in the user network.
\item \textit{Patents}: we can measure the distance between patent categories in the patent citation patterns.
\item \textit{DBpedia}: we do not have node metadata, so we calculate distances between random vectors, but this network could be used, e.g., to calculate distances between different page categories in the encyclopedia, whose pages are connected by hyperlinks.
\end{itemize}

\begin{table*}
\centering
\begin{tabular}{l|rrr|rr|c}
Network & $|V|$ & $|E|$ & Dens & ApproxChol (s) & Baseline (s) & Ref \\
\hline
Hiring & 145 & 2,266 & 0.2170 & 0.0023 & 0.0040 & \cite{clauset2015systematic} \\
EUAir & 450 & 2,953 & 0.0292 & 0.0011 & 0.0454 & \cite{cardillo2013emergence}\\
EUCore & 1,005 & 16,064 & 0.0318 & 0.0068 & 0.4496 & \cite{leskovec2007graph}\\
Open Flights & 3,214 & 18,858 & 0.0036 & 0.0079 & 13.623 & \cite{peixoto2020netzschleuder}\\
LastFm & 7,624 & 27,806 & 0.0009 & 0.0149 & 244.20 & \cite{rozemberczki2020characteristic}\\
Wiki RFA & 11,381 & 194,592 & 0.0030 & 0.0976 & 853.00 & \cite{west2014exploiting}\\
Fly Brain & 21,739 & 2,897,925 & 0.0122 & 2.6151 & 6181.3 & \cite{xu2020connectome}\\
Twitter15m & 87,569 & 4,708,274 & 0.0012 & 4.3299 & & \cite{gonzalez2011dynamics}\\
Patents & 3,774,768 & 16,518,947 & 2.31$e^{-6}$ & 59.311 & & \cite{hall2001nber}\\
DBpedia & 18,268,992 & 136,537,566 & 8.18$e^{-7}$ & 247.10 & & \cite{auer2007dbpedia}\\
\end{tabular}
\caption{The runtimes of the ApproxChol solver against the exact solution in number of seconds for a collection of networks of different sizes (in number of nodes $|V|$ and edges $|E|$). We terminate the process after one hour, thus we do not report runtimes longer than that.}
\label{tab:runtime-real}
\end{table*}

\begin{figure}
\centering
\begin{subfigure}[b]{.33\columnwidth}
\centering
\includegraphics[width=\textwidth]{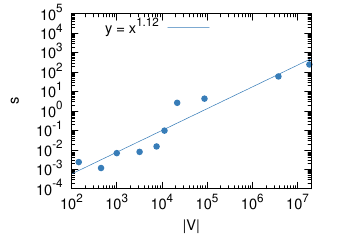}
\caption{ApproxChol}
\end{subfigure}
\qquad
\begin{subfigure}[b]{.33\columnwidth}
\centering
\includegraphics[width=\textwidth]{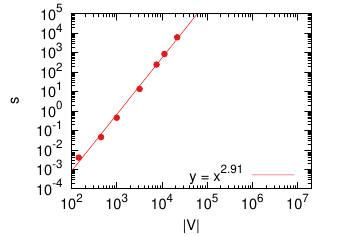}
\caption{Baseline}
\end{subfigure}
\caption{Runtime (y axis) on real networks by number of nodes (x axis).}
\label{fig:runtime-real}
\end{figure}

Table \ref{tab:runtime-real} reports the running times. These are also summarized in Figure \ref{fig:runtime-real}. For comparison purposes, we estimate the scaling of the two methods with a power relation with the number of nodes. The best function approximating the runtime of ApproxChol is $\mathcal{O}(|V|^{1.12})$ (Figure \ref{fig:runtime-real}(a)). On the other hand, the best function approximating the runtime of the exact SVD-based solution is $\mathcal{O}(|V|^{2.91})$ (Figure \ref{fig:runtime-real}(b)).

However, it would be more appropriate to estimate the scaling of the Laplacian solver with the number of edges. This is is because their advantage becomes less and less relevant the more the network is dense. Note the difference in runtimes, e.g., in Hiring and EUAir. Notwithstanding the fact that Hiring has fewer nodes and fewer edges, it is much more dense than EUAir (21\% dense vs 3\% dense) and thus the Laplacian solvers actually take longer to run on this smaller network. On the other hand, it is remarkable that the Laplacian solver can process the DBpedia network (18M nodes) in the same time it takes the baseline to process LastFm (7.6k nodes).

\section{Conclusions}
In this paper we showed that using Laplacian solvers will bring massive speedups in the calculation of the Generalized Euclidean measure and other related measures in the Node Vector Distance class of problems. The speedup is relative to calculating an exact solution via the pseudoinverse of the Laplacian. Since Laplacian solvers scale with the number of edges, these speedups are more noticeable for sparse networks. Besides an improved time efficiency, these methods also require fewer resources in terms of memory, since the process to obtain the pseudoinverse of the Laplacian involves using dense matrices, while all Laplacian solvers work with sparse structures.

We failed to notice significant differences between different Laplacian solvers in synthetic networks. As the network grows in number of nodes, they all increase their runtimes approximately at the same rate. The only potential difference comes when we densify the network. The CG solver is the slowest for very sparse networks, but it scales better as the network becomes denser and denser.

This paper can be used as an argument to use Laplacian solvers to efficiently solve GE and related NVD problems. 

\bibliographystyle{plain}
\bibliography{biblio}

\begin{thebibliography}{10}

\bibitem{andris2015rise}
Clio Andris, David Lee, Marcus~J Hamilton, Mauro Martino, Christian~E Gunning,
  and John~Armistead Selden.
\newblock The rise of partisanship and super-cooperators in the us house of
  representatives.
\newblock {\em PloS one}, 10(4):e0123507, 2015.

\bibitem{auer2007dbpedia}
S{\"o}ren Auer, Christian Bizer, Georgi Kobilarov, Jens Lehmann, Richard
  Cyganiak, and Zachary Ives.
\newblock Dbpedia: A nucleus for a web of open data.
\newblock In {\em The semantic web}, pages 722--735. Springer, 2007.

\bibitem{bajardi2011human}
Paolo Bajardi, Chiara Poletto, Jose~J Ramasco, Michele Tizzoni, Vittoria
  Colizza, and Alessandro Vespignani.
\newblock Human mobility networks, travel restrictions, and the global spread
  of 2009 h1n1 pandemic.
\newblock {\em PloS one}, 6(1):e16591, 2011.

\bibitem{barabasi1999emergence}
Albert-L{\'a}szl{\'o} Barab{\'a}si and R{\'e}ka Albert.
\newblock Emergence of scaling in random networks.
\newblock {\em science}, 286(5439):509--512, 1999.

\bibitem{cardillo2013emergence}
Alessio Cardillo, Jes{\'u}s G{\'o}mez-Gardenes, Massimiliano Zanin, Miguel
  Romance, David Papo, Francisco~del Pozo, and Stefano Boccaletti.
\newblock Emergence of network features from multiplexity.
\newblock {\em Scientific reports}, 3(1):1--6, 2013.

\bibitem{clauset2015systematic}
Aaron Clauset, Samuel Arbesman, and Daniel~B Larremore.
\newblock Systematic inequality and hierarchy in faculty hiring networks.
\newblock {\em Science advances}, 1(1):e1400005, 2015.

\bibitem{cohen2017almost}
Michael~B Cohen, Jonathan Kelner, John Peebles, Richard Peng, Anup~B Rao, Aaron
  Sidford, and Adrian Vladu.
\newblock Almost-linear-time algorithms for markov chains and new spectral
  primitives for directed graphs.
\newblock In {\em Proceedings of the 49th Annual ACM SIGACT Symposium on Theory
  of Computing}, pages 410--419, 2017.

\bibitem{cohen2016faster}
Michael~B Cohen, Jonathan Kelner, John Peebles, Richard Peng, Aaron Sidford,
  and Adrian Vladu.
\newblock Faster algorithms for computing the stationary distribution,
  simulating random walks, and more.
\newblock In {\em 2016 IEEE 57th Annual Symposium on Foundations of Computer
  Science (FOCS)}, pages 583--592. IEEE, 2016.

\bibitem{coifman2006diffusion}
Ronald~R Coifman and St{\'e}phane Lafon.
\newblock Diffusion maps.
\newblock {\em Applied and computational harmonic analysis}, 21(1):5--30, 2006.

\bibitem{coscia2020generalized}
Michele Coscia.
\newblock Generalized euclidean measure to estimate network distances.
\newblock In {\em Proceedings of the International AAAI Conference on Web and
  Social Media}, volume~14, pages 119--129, 2020.

\bibitem{coscia2021atlas}
Michele Coscia.
\newblock The atlas for the aspiring network scientist.
\newblock {\em arXiv preprint arXiv:2101.00863}, 2021.

\bibitem{coscia2021pearson}
Michele Coscia.
\newblock Pearson correlations on complex networks.
\newblock {\em Journal of Complex Networks}, 9(6):cnab036, 2021.

\bibitem{coscia2020node}
Michele Coscia, Andres Gomez-Lievano, James Mcnerney, and Frank Neffke.
\newblock The node vector distance problem in complex networks.
\newblock {\em ACM Computing Surveys (CSUR)}, 53(6):1--27, 2020.

\bibitem{coscia2022posts}
Michele Coscia and Clara Vandeweerdt.
\newblock Posts on central websites need less originality to be noticed.
\newblock {\em Scientific reports}, 12(1):1--10, 2022.

\bibitem{devriendt2022variance}
Karel Devriendt, Samuel Martin-Gutierrez, and Renaud Lambiotte.
\newblock Variance and covariance of distributions on graphs.
\newblock {\em SIAM Review}, 64(2):343--359, 2022.

\bibitem{friggeri2014rumor}
Adrien Friggeri, Lada Adamic, Dean Eckles, and Justin Cheng.
\newblock Rumor cascades.
\newblock In {\em proceedings of the international AAAI conference on web and
  social media}, volume~8, pages 101--110, 2014.

\bibitem{ghasemian2020stacking}
Amir Ghasemian, Homa Hosseinmardi, Aram Galstyan, Edoardo~M Airoldi, and Aaron
  Clauset.
\newblock Stacking models for nearly optimal link prediction in complex
  networks.
\newblock {\em Proceedings of the National Academy of Sciences},
  117(38):23393--23400, 2020.

\bibitem{gonzalez2011dynamics}
Sandra Gonz{\'a}lez-Bail{\'o}n, Javier Borge-Holthoefer, Alejandro Rivero, and
  Yamir Moreno.
\newblock The dynamics of protest recruitment through an online network.
\newblock {\em Scientific reports}, 1(1):1--7, 2011.

\bibitem{granovetter1978threshold}
Mark Granovetter.
\newblock Threshold models of collective behavior.
\newblock {\em American journal of sociology}, 83(6):1420--1443, 1978.

\bibitem{hall2001nber}
Bronwyn~H Hall, Adam~B Jaffe, and Manuel Trajtenberg.
\newblock The nber patent citation data file: Lessons, insights and
  methodological tools, 2001.

\bibitem{hausmann2014atlas}
Ricardo Hausmann, C{\'e}sar~A Hidalgo, Sebasti{\'a}n Bustos, Michele Coscia,
  and Alexander Simoes.
\newblock {\em The atlas of economic complexity: Mapping paths to prosperity}.
\newblock Mit Press, 2014.

\bibitem{hohmann2023quantifying}
Marilena Hohmann, Karel Devriendt, and Michele Coscia.
\newblock Quantifying ideological polarization on a network using generalized
  euclidean distance.
\newblock {\em Science Advances}, 9(9):eabq2044, 2023.

\bibitem{holland1983stochastic}
Paul~W Holland, Kathryn~Blackmond Laskey, and Samuel Leinhardt.
\newblock Stochastic blockmodels: First steps.
\newblock {\em Social networks}, 5(2):109--137, 1983.

\bibitem{koutis2011nearly}
Ioannis Koutis, Gary~L Miller, and Richard Peng.
\newblock A nearly-m log n time solver for sdd linear systems.
\newblock In {\em 2011 IEEE 52nd Annual Symposium on Foundations of Computer
  Science}, pages 590--598. IEEE, 2011.

\bibitem{koutis2014approaching}
Ioannis Koutis, Gary~L Miller, and Richard Peng.
\newblock Approaching optimality for solving sdd linear systems.
\newblock {\em SIAM Journal on Computing}, 43(1):337--354, 2014.

\bibitem{kraemer2020effect}
Moritz~UG Kraemer, Chia-Hung Yang, Bernardo Gutierrez, Chieh-Hsi Wu, Brennan
  Klein, David~M Pigott, Open COVID-19 Data~Working Group†, Louis Du~Plessis,
  Nuno~R Faria, Ruoran Li, et~al.
\newblock The effect of human mobility and control measures on the covid-19
  epidemic in china.
\newblock {\em Science}, 368(6490):493--497, 2020.

\bibitem{kyng2016approximate}
Rasmus Kyng and Sushant Sachdeva.
\newblock Approximate gaussian elimination for laplacians-fast, sparse, and
  simple.
\newblock In {\em 2016 IEEE 57th Annual Symposium on Foundations of Computer
  Science (FOCS)}, pages 573--582. IEEE, 2016.

\bibitem{leskovec2007dynamics}
Jure Leskovec, Lada~A Adamic, and Bernardo~A Huberman.
\newblock The dynamics of viral marketing.
\newblock {\em ACM Transactions on the Web (TWEB)}, 1(1):5--es, 2007.

\bibitem{leskovec2007graph}
Jure Leskovec, Jon Kleinberg, and Christos Faloutsos.
\newblock Graph evolution: Densification and shrinking diameters.
\newblock {\em ACM transactions on Knowledge Discovery from Data (TKDD)},
  1(1):2--es, 2007.

\bibitem{snapnets}
Jure Leskovec and Andrej Krevl.
\newblock {SNAP Datasets}: {Stanford} large network dataset collection.
\newblock \url{http://snap.stanford.edu/data}, June 2014.

\bibitem{lewis2019voteview}
Jeffrey~B Lewis, Keith Poole, Howard Rosenthal, Adam Boche, Aaron Rudkin, and
  Luke Sonnet.
\newblock Voteview: Congressional roll-call votes database.
\newblock {\em https://voteview. com/ (accessed 25 February 2022)}, 2019.

\bibitem{peixoto2020netzschleuder}
Tiago~P Peixoto.
\newblock The netzschleuder network catalogue and repository, 2020.

\bibitem{pele2009fast}
Ofir Pele and Michael Werman.
\newblock Fast and robust earth mover's distances.
\newblock In {\em 2009 IEEE 12th international conference on computer vision},
  pages 460--467. IEEE, 2009.

\bibitem{rozemberczki2020characteristic}
Benedek Rozemberczki and Rik Sarkar.
\newblock Characteristic functions on graphs: Birds of a feather, from
  statistical descriptors to parametric models.
\newblock In {\em Proceedings of the 29th ACM international conference on
  information \& knowledge management}, pages 1325--1334, 2020.

\bibitem{shuman2013emerging}
David~I Shuman, Sunil~K Narang, Pascal Frossard, Antonio Ortega, and Pierre
  Vandergheynst.
\newblock The emerging field of signal processing on graphs: Extending
  high-dimensional data analysis to networks and other irregular domains.
\newblock {\em IEEE signal processing magazine}, 30(3):83--98, 2013.

\bibitem{shuman2016vertex}
David~I Shuman, Benjamin Ricaud, and Pierre Vandergheynst.
\newblock Vertex-frequency analysis on graphs.
\newblock {\em Applied and Computational Harmonic Analysis}, 40(2):260--291,
  2016.

\bibitem{spielman2004nearly}
Daniel~A Spielman and Shang-Hua Teng.
\newblock Nearly-linear time algorithms for graph partitioning, graph
  sparsification, and solving linear systems.
\newblock In {\em Proceedings of the thirty-sixth annual ACM symposium on
  Theory of computing}, pages 81--90, 2004.

\bibitem{spielman2011spectral}
Daniel~A Spielman and Shang-Hua Teng.
\newblock Spectral sparsification of graphs.
\newblock {\em SIAM Journal on Computing}, 40(4):981--1025, 2011.

\bibitem{spielman2013local}
Daniel~A Spielman and Shang-Hua Teng.
\newblock A local clustering algorithm for massive graphs and its application
  to nearly linear time graph partitioning.
\newblock {\em SIAM Journal on computing}, 42(1):1--26, 2013.

\bibitem{spielman2014nearly}
Daniel~A Spielman and Shang-Hua Teng.
\newblock Nearly linear time algorithms for preconditioning and solving
  symmetric, diagonally dominant linear systems.
\newblock {\em SIAM Journal on Matrix Analysis and Applications},
  35(3):835--885, 2014.

\bibitem{tacchella2012new}
Andrea Tacchella, Matthieu Cristelli, Guido Caldarelli, Andrea Gabrielli, and
  Luciano Pietronero.
\newblock A new metrics for countries' fitness and products' complexity.
\newblock {\em Scientific reports}, 2(1):1--7, 2012.

\bibitem{vespignani2020modelling}
Alessandro Vespignani, Huaiyu Tian, Christopher Dye, James~O Lloyd-Smith,
  Rosalind~M Eggo, Munik Shrestha, Samuel~V Scarpino, Bernardo Gutierrez,
  Moritz~UG Kraemer, Joseph Wu, et~al.
\newblock Modelling covid-19.
\newblock {\em Nature Reviews Physics}, 2(6):279--281, 2020.

\bibitem{vishnoi2013lx}
Nisheeth~K Vishnoi et~al.
\newblock Lx= b.
\newblock {\em Foundations and Trends{\textregistered} in Theoretical Computer
  Science}, 8(1--2):1--141, 2013.

\bibitem{von2007tutorial}
Ulrike Von~Luxburg.
\newblock A tutorial on spectral clustering.
\newblock {\em Statistics and computing}, 17:395--416, 2007.

\bibitem{watts2007viral}
Duncan~J Watts, Jonah Peretti, and Michael Frumin.
\newblock {\em Viral marketing for the real world}.
\newblock Harvard Business School Pub. Boston, 2007.

\bibitem{watts1998collective}
Duncan~J Watts and Steven~H Strogatz.
\newblock Collective dynamics of ‘small-world’networks.
\newblock {\em nature}, 393(6684):440--442, 1998.

\bibitem{weng2014predicting}
Lilian Weng, Filippo Menczer, and Yong-Yeol Ahn.
\newblock Predicting successful memes using network and community structure.
\newblock In {\em Eighth international AAAI conference on weblogs and social
  media}, 2014.

\bibitem{west2014exploiting}
Robert West, Hristo~S Paskov, Jure Leskovec, and Christopher Potts.
\newblock Exploiting social network structure for person-to-person sentiment
  analysis.
\newblock {\em Transactions of the Association for Computational Linguistics},
  2:297--310, 2014.

\bibitem{xu2020connectome}
C~Shan Xu, Michal Januszewski, Zhiyuan Lu, Shin-ya Takemura, Kenneth~J
  Hayworth, Gary Huang, Kazunori Shinomiya, Jeremy Maitin-Shepard, David
  Ackerman, Stuart Berg, et~al.
\newblock A connectome of the adult drosophila central brain.
\newblock {\em BioRxiv}, 2020.

\bibitem{zhao2019moverscore}
Wei Zhao, Maxime Peyrard, Fei Liu, Yang Gao, Christian~M Meyer, and Steffen
  Eger.
\newblock Moverscore: Text generation evaluating with contextualized embeddings
  and earth mover distance.
\newblock {\em arXiv preprint arXiv:1909.02622}, 2019.

\end{thebibliography}

\end{document}